\documentstyle[12pt]{article}
\newcommand{\be}{\begin{equation}}
\newcommand{\ee}{\end{equation}}
\newcommand{\bea}{\begin{eqnarray}}
\newcommand{\eea}{\end{eqnarray}}

\begin{document}
\begin{center}
{\Large\bf The Gluon Propagator in Analytic Perturbation Theory \\}

 \medskip

{\bf B.A. Magradze\footnote {E-mail address:
magr@rmi.acnet.ge}}
\medskip

{\it Tbilisi Mathematical Institute,
Georgian Academy of Sciences,\\
Tbilisi 380093
Georgia\\}

\medskip

{\footnotesize \bf Abstract}
\medskip

\parbox{110mm}{\footnotesize

The structure of the  $\beta$-function of massless QCD is considered
in analytic perturbation theory (APT).
The two-loop analytic effective coupling constant is written in terms of the 
Lambert W function.
The method  of APT is applied to the gluon propagator in the Landau gauge.
It is shown that there is an nonperturbative ambiguity in determination
of the anomalous dimension function of the gluon field.
One possible resolution of the ambiguity is presented.}

\end{center}

           \bigskip

{\bf 1.Introduction}\\
\medskip

	Consistent theoretical determination  of the effective coupling
constant  has become one of the most important tasks
of  QCD. A serious difficulty is the occurrence of  Landau singularities
in the  effective coupling constant in perturbation theory (PT).
It is widely believed that in realistic quantum field theories these
unphysical singularities are artifact of  PT
\cite{red,bog}.
In QCD they  appear  at small space-like momenta.
For this reason PT  becomes useless for low-energy calculations.
Supposedly, these  infrared singularities signal that the
perturbative vacuum differs from the true vacuum of the theory \cite{mand}.
In recent years more and more attention has been focused on this problem 
in connection with large order behavior of QCD perturbative expansions.   

	On the other hand, most of  the suggestions of the confinement
mechanism
include  assumptions about  the infrared behaviour of the effective coupling
constant. Thus in the
popular linear confinement picture  singular infrared behaviour is assumed
for the effective coupling constant $\bar \alpha_{s}(q^2) \sim q^{-2}$.
It was confirmed that this behaviour is consistent with
the  Schwinger-Dyson equations (SDEs) of QCD \cite{mand,bak}.

	Independently, in the literature  the hypothesis
about infrared finite effective coupling constant \cite{bank} 
has become popular.
Several convincing arguments have been 
proposed to support the idea that the running coupling "freezes" at low
energies \cite{stev}. The confinement mechanism which may be consistent with
this
hypothesis
has been proposed in Refs. \cite{nishi}.  A linearly rising potential is not
assumed in this framework.

One   possible resolution of the "Landau-ghost" problem  has been
suggested
long time ago in Refs. \cite{red,bog}, where the
"analytization procedure" was elaborated in PT.
Recently, this method has been  successfully applied to the QCD
running coupling \cite{shir1,shir2,sol1,sol2,sol3,sol4,my}.
An Important observation has been made
in Refs.  \cite{shir1}.
It was found that the modified solution for $\bar\alpha_{s}(-q^{2})$ has
a finite infrared  limit  $\bar\alpha_{s}(0)$, which is
stable with respect to higher order corrections.
The method has been applied to the inclusive decay of a $\tau$ lepton
\cite{sol2}, and it was shown that the renormalization scheme ambiguity of
the results is considerably reduced in APT \cite{sol3,sol4}. Furthermore,
APT provides a self-consistent determination of the running coupling constant
in the time-like region \cite{sol1}.
An approach similar in spirit has been followed by the authors of Refs.
\cite{gr,dk}.

The organization of the paper is as follows. In Sec.2  we present
the notation and discuss some technical aspects of analytic perturbation theory
(APT). The structure of the $\beta$-function is studied in APT.
We write the  two-loop analytic effective
coupling constant   in terms of Lambert W function
\cite{lamb}.
Thus we avoid
Newton's iteration method of solution.
The   iterative solution is  compared numerically with the exact solution
in the infrared domain.
In Sec.3 we apply the method to the gluon propagator in the Landau gauge.
We have found that the K\"{a}llen-Lehmann analyticity and the RG
alone are not sufficient to uniquely   determine
the nonperturbative part of the anomalous dimension function for the gluon
field. Two different  definitions for  the anomalous dimension
function are considered. We discuss a possible resolution of the ambiguity.
Sec. 4 contains the concluding remarks.
\bigskip

{\bf 2. The "analytization procedure" for the QCD running coupling}\\

The $ \beta $-function in QCD is given by
\be \label{beta}
{\mu}^2\frac{d{\alpha}_{s}}{d{\mu}^2}={\beta}({\alpha}_{s}),
\ee
where ${\alpha}_{s}=\frac{g^2}{4{\pi}} $,  $ g^2 $ is the renormalized
coupling constant, and $\mu$ is the renormalization point. For  simplicity,
we have assumed that the quarks are massless. We have in mind some
renormalization scheme, which we do not need to specify at this stage.
For convenience we choose the covariant Landau
gauge. Eq.~(\ref{beta}) defines the $\beta$-function order by order in PT.
However,
one  can assume Eq.~(\ref{beta})  beyond of PT. The assumption is that
the non trivial $\beta$ function exists with perturbative
asymptotic expansion in powers of ${\alpha}_{s}$
\be
\label{beta1}
\beta({\alpha}_{s})\approx-\frac{\beta_0}{4\pi}({\alpha}_{s})^2
-\frac{\beta_1}{(4\pi)^2}({\alpha}_{s})^3
+\cdots
\ee
The first two coefficients of the series are independent of the chosen
renormalization scheme. Their values are given by
\be
\label{coef}
\beta_{0}=(11-2N_{f}/3),\hspace{5mm}
\beta_{1}=(102-38N_{f}/3),
\ee
with $N_{f}$ being the number of flavors.
The effective coupling constant is defined by
\be
\label{eff}
k^2\frac{d}{dk^2}{\bar{\alpha}_{s} }(-k^2)=\beta({\bar{\alpha}_{s} }(-k^2)),\ee
\be
\label{norm}
{\bar{\alpha}_{s} }({\mu}^2)={\alpha}_{s}.
\ee
Convenient definition of the QCD scale parameter $\Lambda$ has been given in
Ref.\cite{stev1}. The relation between conventionally defined
parameter $\Lambda_{cn}$ and
$\Lambda$ is given by $\Lambda_{cn}=(\eta)^{(\eta/2)}\Lambda$ with
$\eta=\beta_{1}/\beta_{0}^2$.
With this definition and with normalization
(\ref{norm}) the solution of  Eq.~(\ref{eff})
is
\be
\label{pt}
\ln\left(\frac{-k^{2}}{{\Lambda}^2}\right)=
\frac{4\pi}{\beta_{0}{\bar\alpha}_{s} }-\frac{\beta_1}{\beta_0^2}\ln\left(1+
\frac{4\pi\beta_0}{\beta_1{\bar\alpha}_{s}}\right)
+\psi(\bar{\alpha}_{s}),
\ee
where
$$
\psi({\alpha}_{s})= \int_{0}^{{\alpha}_{s}}(1/\beta(x)
-1/\beta^{(2)}(x))dx,
$$
and ${\beta}^{(2)}(x)$ denotes the two-loop $\beta$-function.
To express $\bar\alpha_{s}(-k^2)$ explicitly as a function of $k^2$,
beyond the one-loop order, one must solve the transcendental equation (\ref{pt}).
Usually, iteration procedure is used.

Let $\bar\alpha_{pt}^{(n)}(-k^2)$ be  nth-order perturbative
solution for $\bar\alpha_{s}$ determined by Eq.~(\ref{pt}).
Then the corresponding analytic solution is given by the K\"{a}llen-Lehmann integral
\be
\label{kl}
\bar\alpha_{an}^{(n)}(-k^2) =\frac{1}{\pi}\int_{0}^{\infty}
\frac{{\rho}^{(n)}(\sigma)}{(\sigma-k^2-i0)}d\sigma,
\ee
where spectral function $ {\rho}^{(n)}(\sigma) $ is the discontinuity of
$\bar\alpha_{pt}^{(n)}(-k^2)$ along the positive $k^2$ axis.
The main result of Refs.\cite{shir1} is that the value $\bar\alpha_{(an)}(0) $
in APT is determined by one-loop calculation:
$$
\bar\alpha_{an}^{(n)}(0)=\bar\alpha_{an}^{(1)}(0)=4\pi/\beta_{0}
=4\pi/(11-2N_{f}/3).
$$

Using Cauchy's theorem  one can represent the function (\ref{kl}) in the equivalent
form
\be
\label{anco}
\bar\alpha_{an}^{(n)}(-k^2)=\bar\alpha_{pt}^{(n)}(-k^2)+
{\theta}^{(n)}(-k^2),
\ee
the "nonperturbative"
term ${\theta}^{(n)}$  compensates the unphysical
contributions
in  $\bar\alpha_{pt}^{(n)}$. We shall assume that ${\theta}^{(n)}$
satisfies the
representation
\be
\label{theta}
{\theta}^{(n)}(-k^2)=
\frac{1}{\pi}\int_{k_{L}^2}^{0}
\frac{{\rho}_{\theta}^{(n)}(\sigma)}{(\sigma-k^2-i0)}d\sigma,
\ee
where $k_{L}^2=-c_{(n)}{\Lambda}^2$, and $c_{(n)}$ is  the positive real
number calculable
in PT, for example
$c_{(1)}=c_{(2)}=1$.   To third order, in the $MS$ scheme and for $N_{f}=3$,
we find that $c_{(3)}=1.87$ \footnote{Note that the one-iteration solution
of Eq.(\ref{pt}) develops unphysical complex singularities for $n\ge3$.}.
The normalization condition (\ref{norm}) now reads
\be
\label{norm1}
\bar\alpha_{an}({\mu}^2)=\bar\alpha_{pt}^{(n)}({\mu}^2)+{\theta}^{(n)}
({\mu}^2)=\alpha_{s}.
\ee
Let us consider the structure of the $\beta$-function in APT.
From  Eqs.~(\ref{kl}) and (\ref{norm1}) we have
\be
\label{betn}
\beta_{an}^{(n)}(\alpha_{s})=-\frac{1}{\pi}\int_{0}^{\infty}
\frac{{\mu}^2}{(\sigma+{\mu}^2)^2}\rho^{(n)}(\sigma)d\sigma=
-\frac{1}{\pi}\int_{0}^{\infty}\frac{\bar\rho^{(n)}(s)}{\xi(s+1/\xi)^2}ds,
\ee
where $\bar\rho^{(n)}(s)=\rho^{(n)}({\Lambda}^{2}s)$ and $\xi$ denotes
the dimensionless variable
$
\xi=exp(-\phi)={\Lambda}^2/{\mu}^2.
$
Note that $\xi\sim\exp(-4\pi/\beta_{0}\alpha_{s})$ for
$\alpha_{s}\rightarrow 0$. From Eq.(\ref{betn}) we see that
$\beta(\alpha_{s})\equiv\hat{\beta}(\xi)$ is analytical in the cut-$\xi$ plane.
Alternatively, from Eq.(\ref{norm1})  we find
\be
\label{betan}
{\beta}^{(n)}_{an}(\alpha_{s})=\partial \alpha_{s}/ \partial ln \mu^2=
{\beta}^{(n)}_{pt}(\bar\alpha_{pt}^{(n)}(\mu^2))+\bar{\theta}^{(n)}(\mu^2)=
{\beta}^{(n)}_{pt}(\alpha_{s}-{\theta}^{(n)}(\mu^2))+
\bar{\theta}^{(n)}(\mu^2),
\ee
where ${\beta}^{(n)}_{pt}(x)=-\sum_{k=0}^{n-1}\beta_{k}x^{k+2}/(4\pi)^{k+1},$
and $\bar{\theta}^{(n)}({\mu}^2)={\mu}^2\frac{\partial {\theta}^{(n)}({\mu}^2)}
{\partial {\mu}^2}.$
We can rewrite Eq(\ref{betan})
in the form
\be
\label{betan1}
{\beta}_{an}({\alpha}_{s})={\beta}_{pt}({\alpha}_{s})+{\beta}_{np}
({\alpha}_{s},\theta,\bar\theta),
\ee
here, for convenience we have used condensed notations suppressing the
superscript (n) and arguments  to the expressions above
($\theta\equiv{\theta}^{(n)}({\mu}^2)$ ets.).
The nonperturbative piece  ${\beta}_{np}$   can be written as a power
series for $ {\alpha}_{s} $
\be
\label{np1}
{\beta}_{np}({\alpha}_{s},\theta,\bar\theta)=\sum_{k=0}^{n}
B_{k}(\theta,\bar\theta)(\alpha_{s})^{k},
\ee
where
\be
\label{npc}
B_{0}=\bar\theta,\hspace{3mm}
B_{1}=-\frac{\partial\beta_{pt}(-\theta)}
{\partial\theta},
\hspace{3mm}
B_{k}(\theta)=\frac{(-1)^k}{k!}\frac{{\partial}^{k}{{\beta}_{pt}}(-\theta)}
{\partial\theta^{k}}+\frac{\beta_{k-2}}{(4\pi)^{k+1}} \hspace{3mm} for \hspace{3mm} k\ge 2,
\ee
From the representation (\ref{theta})  we see that $\theta({\mu}^2)$
and $\bar\theta({\mu}^2)$ are analytical functions in the cut $\xi$-plane.
The cut is along the positive real $\xi$-axis $\xi\ge 1/c_{(n)}$.
Then,  from (\ref{npc}) it follows that the coefficients
$B_{k}(\theta)\equiv\hat{B}_{k}(\xi)$ are analytical in the domain
$\xi\le 1/c_{(n)}$. In the weak-coupling limit
they vanish
exponentially $\hat{B}_{k}(\xi)\sim \xi\sim\exp(-\frac{4\pi}
{\beta_{0}\alpha_{s}})$.
A detailed study of the one-loop $\beta$-function can be found in Ref.~
\cite{shir2}.

Now, let us consider the two-loop running coupling in APT more detailed.
It is convenient to introduce  quantities
\be
\label{newdef}
\bar a_{s}(-k^2)=\frac{\beta_{0}}{4\pi}\bar\alpha_{s}(-k^2) \hspace{1cm}
and\hspace{1cm}
\bar\beta(a_{s})=\frac{\partial a_{s}}{\partial \ln({\mu}^2)}
\ee
To second order Eq.~(\ref{pt})  reads
\be
\label{trans}
\ln(z)=\frac{1}{y(z)}-\eta \ln\left(1+\frac{1}{\eta}\frac{1}{y(z)}
\right),
\ee
where we denote $z=-k^2/{\Lambda}^2$, $y(z)=\bar a_{s}^{(2)}(-k^2)$ and
$\eta=\beta_1/{\beta_0}^2$
$(\eta=64/81$ for $ N_{f}=3)$. We shall accept the flavor condition
$N_{f}<8.05$, i.e. $\beta_{1}>0$. In this case the Landau singularity is
occurred in the two-loop running coupling \cite{bank,stev}.
The transcendental equation (\ref{trans}) is solved   by the iteration
method.
The solution given by one iteration is
\be
\label{iter}
y_{it}(z)=\frac{1}{\ln(z)+\eta \ln\left(1+\frac{1}{\eta}\ln(z)\right)}.
\ee
The function (\ref{iter}) provides good  approximation,  for a small
enough value of $\bar a_{s} $.
Consider (\ref{iter}) as an analytic function in the whole
complex z-plane. It has the Landau pole at $z=1$, the logarithmic branch point at
$z=exp(-\eta)$ and standard branch point at $z=0$.
The corresponding analytic solution is given by the K\"{a}llen-Lehmann integral
(see Refs.\cite{shir1})
\be
\label{kl1}
y_{it,an}(z)=\frac{1}{\pi}\int_{0}^{\infty}\frac{\bar\rho_{it}^{(2)}(x)}{x+z}dx
=\frac{1}{\pi}\int_{-\infty}^{\infty}\frac{e^{t}}{e^{t}+z}\tilde
\rho_{it}^{(2)}(t) dt,
\ee
where $\tilde\rho^{(2)}_{it}(t)\equiv\bar\rho^{(2)}_{it}(e^t)\equiv
\rho^{(2)}_{it}(e^{t}\Lambda^{2})=L_{1}/(L_{1}^{2}+L_{2}^{2})$ with
$$
L_{1}=\pi+\eta \arccos((\eta+t)/r),\hspace{0.3cm}L_2=t+\eta \ln(r/\eta),\hspace{0.3cm} r=\sqrt{(\eta+t)^2+\pi^2}.
$$
Using the Lagrange
Inversion Formula
we see from (\ref{trans}) that z=1  is not a pole but the second order brunch
point:
$y(z)\sim \frac{1}{\sqrt{2\eta(z-1)}}$ for $z\rightarrow 1$.
Thus, (\ref{iter}) violates the analytical properties of $y(z)$ near the point
z=1.
Therefore, it is  not unreasonable to solve  exactly Eq.~(\ref{trans})
and to estimate accuracy of  the solution (\ref{kl1}).
Note that the transcendental Eq.~(\ref{trans}) is exactly solvable.
Indeed, it is easy to convince that the solution is
\be
\label{w}
y(z)=-\frac{1}{\eta}\frac{1}{1+W(\zeta)},
\ee
where
\be
\label{zeta}
\zeta=-\frac{1}{e}z^{-\frac{1}{\eta}},
\ee
and $W$ denotes the multivalued inverse of $\zeta=W\exp(W)$. That is the
Lambert W function. A detailed review of the properties and applications
of this interesting  function can be found in \cite{lamb}.
In this article we use definition and notation of branches of $W$
according  the computer algebra system Maple Release 2
\footnote{In Ref.[13] different conventions are accepted.}.
Now our task is to choose the suitable branches of $W$ in Eq.~(\ref{w}).
The solution must be regular in the cut complex z-plane.
Consider the function (\ref{zeta}) for $0<\eta< 1$.
It maps the cut  z-plane
onto the three- sheet  Riemannian surface of $\zeta$.
Combining above consideration with known properties of the branches of
$W$ we find
\be
\label{w1}
W(\zeta)=\left\{\begin{array}{ll}
                  W(0,\zeta) &\mbox{if $|arg(z)|\leq \eta\pi $}\\
                  W(-1,\zeta) &\mbox{if $\eta\pi\leq arg(z) \leq\pi $}\\
                  W(1,\zeta) &\mbox{if $-\pi\leq arg(z)\leq-\eta\pi $}
                  \end{array}
                  \right.
\ee
Taking into account Eqs.~(\ref{kl}), (\ref{w}) and (\ref{w1}),
we write the K\"{a}llen-Lehmann representation
for the corresponding causal solution
\be
\label{kl2}
y_{an}(z)=\frac{1}{\pi}\int_{0}^{\infty}\frac{\bar\rho^{(2)}(x)}{x+z}dx=
\frac{1}{\pi}\int_{-\infty}^{\infty}\frac{e^{t}}{(e^{t}+z)}{\tilde\rho^{(2)}(t)}dt,
\ee
where $\tilde\rho^{(2)}(t)\equiv\bar\rho^{(2)}(e^t)\equiv\rho^{(2)}(e^{t}
\Lambda^{2})$
and
\be
\label{ro}
\tilde\rho^{(2)}(t)=-\frac{1}{\eta}Im\left(\frac{1}{1+W(1,\zeta_{1}(t))}\right),
\hspace{0.5cm}\zeta_{1}(t)=\exp\left(-\frac{t}{\eta}-1+i(\frac{1}{\eta}-1)
\pi\right).
\ee
Equivalently, using the Cauchy formula we can rewrite Eq.~(\ref{kl2}),
for $0\leq z \leq 1$,
as follows
$$
y_{an}(z)=-\frac{1}{\eta}Re\left(\frac{1}{1+W(0,-\frac{1}{ez^{\frac{1}{\eta}}})}
\right)+\frac{1}{\eta\pi}PV \int_{0}^{1} \frac{1}{x-z}Im\left(\frac{1}
{1+W(0,-\frac{1}{ex^{\frac{1}{\eta}}})}\right)dx,
$$
where PV denotes the principal value of the integral along the real axis from 0
to 1.
Numerical results for the exact two-loop function (\ref{kl2}) as well as for the
iterative solution (\ref{kl1}) are summarized in the Table 1.
\begin{table}[h]
\begin{tabular}{|l|l|l|c|l|l|l|}\hline
$-k^2/{{\Lambda}^2} $   & $ \bar a_{an}^{(2)}(- k^2)$ & 
$ \bar a_{an}^{(2).it}(- k^2)$
& &
${-k^2}/{{\Lambda}^2} $   & $ \bar a_{an}^{(2)}(-k^2)$ & 
$\bar a_{an}^{(2).it}(-k^2)$ \\
\hline
0   & 1.000     & 1.000  &    & 6   & 0.252    & 0.261  \\
0.50 & 0.397    & 0.405  &    & 7   & 0.245    & 0.253  \\
1    & 0.352    & 0.362  &    & 8   & 0.238    & 0.247  \\
2    & 0.310    & 0.320  &    & 9   & 0.233    & 0.241  \\
3    & 0.287    & 0.297  &    & 10  & 0.229    & 0.237    \\
4    & 0.272    & 0.281  &    & 15  & 0.212    & 0.219     \\
5    & 0.261    & 0.270  &    & 20  & 0.201   & 0.208          \\ \hline
\end{tabular}
\caption{One-iteration approximation vs exact solution for $N_{f}=3$}
\end{table}
The relative error for the solution (\ref{kl1}),
is about 3\% for the considered
interval. This gives 8.3\% error in the value of the QCD scale parameter
$\Lambda^{(2)}$ for $N_{f}=3$ (see Table 2). Following Refs.\cite{shir1} we
may use the
average
$$
A(Q)=\frac{1}{Q}\int_{0}^{Q}\bar{\alpha}_{s}(\mu^2)d\mu
$$
for comparison. For $ Q=2 $ GeV  we find that iterative solution
(\ref{kl1})
gives answer to much better than $ 0.5\% $ accuracy. The reference values of
$ \alpha_{an}(M_{\tau}) $ are taken from Refs.\cite{shir1}:
$ \alpha_{an}(M_{\tau})=0.36 \pm 0.02 $ with  $ M_{\tau}=1.78$ GeV.
\bigskip
\begin{table}
\begin{tabular}{|l|l|l|l|}\hline
$ \alpha_{an}(M_{\tau}) $& 0.34 & 0.36 &0.38 \\ \hline
$\Lambda^{(2)}_{it}$(MeV) & 610 & 710 & 820\\
$\Lambda^{(2)}$(Mev) & 670 & 770 & 890\\
$A^{it}(2GeV)$& 0.477 & 0.500 & 0.523\\
$A(2GeV)$& 0.480  & 0.502 & 0.525 \\ \hline
\end{tabular}
\caption{}
\end{table}
\bigskip
It should be noted that in the recent article \cite{sol4}  the
transcendental Eq.(\ref{trans}) for the running coupling constant
has been solved numerically on the physical cut.
Nevertheless, exact expression (\ref{kl2}) is itself  valuable if we
have in mind to explore global analytical properties of the solution.

The procedure is not unique. Indeed, one can redefine the analytic charge
according
$$ \bar \alpha_{an}^{new}(-k^2)=\bar \alpha_{an}(-k^2) +
\delta\bar \alpha_{an}(-k^2) \hspace{3mm}with\hspace{3mm}
\delta\bar \alpha_{an}(-k^2)=
\frac{1}{\pi}\int\limits_{0}^{K^2(\alpha_{s},\mu)}
\frac{\omega(\sigma)}{\sigma-k^2-i0}d\sigma
$$
where $\delta\bar \alpha_{an}$ is a "genuine" nonperturbative
correction, which cannot be reproduced in APT, $\omega(\sigma)$ is a
integrable function and $K(\alpha_{s},\mu)$ is invariant under the RG.
New analytic charge is asymptotically free and in the weak limit
it reproduces correct perturbative series. In principle one can
fix this ambiguity using appropriate truncated set of the
SDEs. Indeed,
in the recent publications \cite{aa} reasons have given to believe that
$\delta\bar\alpha_{an}$ cannot be  zero, and an infrared regular QCD running
coupling constant cannot be consistent with the SDEs
for the gluon propagator. Ghost-free axial gauge has been chosen in this work.
On the contrary, in Ref.\cite{sha} the infrared finite effective coupling
constant
has been realized within the framework of the SDEs  in the Landau gauge.
However, it was shown that the solution for the gluon propagator does
not satisfy the K\"{a}llen-Lehmann analyticity. We see that situation is not so
clear.

Nevertheless, APT as it was formulated in Refs.\cite{red,bog}
has advantages. It is a minimal causal extension of PT and
there are no extra parameters in addition to $\alpha_{s}$. Another attractive
feature is the stability of the results in  APT \cite{shir1}.
APT may provide a valuable approximation in the infrared domain
where conventional PT fails.
\bigskip

{\bf 3. The gluon propagator in APT}\\

The method  discussed above originally  has been presented
for calculating modified propagators within framework of PT \cite{red}.
However, the author  of \cite{red} did not explore renormalization invariance
of the theory. This step was taken in paper \cite{bog}, where the procedure
has been formulated for the effective coupling constants in renormalizable
theories. The method was used for the photon propagator.
In QED  due to the Ward identity simple relation holds
\be
\label{gam}
{\gamma}_{v}(\alpha)=\beta(\alpha)/\alpha,
\ee
here $\gamma_{v}(\alpha)$ is the anomalous dimension of the gauge field.
Because of (\ref{gam}) the QED effective coupling is proportional to the
transverse
dimensionless structure of the photon propagator.
In QCD the relation (\ref{gam})  generally is not valid
\footnote{  (\ref{gam})  is still valid in axial-type gauges, as well as,
in the background field gauge.}.
Therefore, the method can not be applied  to the gluon
propagator unambiguously .
Let us consider the gluon propagator in the Landau gauge
$
D_{\mu\nu}(k)=(g_{\mu\nu}-k_{\mu}k_{\nu}/k^2)D(-k^2).
$
We assume that quarks are massless and normalize the propagator at the
Euclidean point $k^2=-{\mu}^2$
\be
\label{gl}
D(-k^2)=-d(-k^2/{\mu}^2,\alpha_{s})/k^2,\hspace{0.5cm}d(1,g^2)=1.
\ee
 Invariance under the RG, in the Landau gauge, leads
\be
\label{gs}
d(-k^2/\mu^2, \alpha_{s})=\exp\left(\int_{\alpha_{s}}^
{\bar \alpha_{s}
(-k^2)}\frac{\gamma_{v}(x)}
{\beta(x)}dx\right)
\ee
The formal power series for the anomalous dimension function $\gamma_{v}$,
in the Landau gauge, is given by
\be
\label{gv}
{\gamma}_{v}(\alpha_{s})=-(\frac{{\gamma_{0}}}{4\pi}\alpha_{s}+
\frac{\gamma_{1}}{16{\pi}^2}{\alpha_{s}}^2
+\ldots),
\ee
with ${\gamma}_{0}=(13-4N_{f}/3)/2$ and
$\gamma_{1}=(531/8-61N_{f}/6)$.

 A detailed investigation of the gluon propagator in the framework of
 analyticity and asymptotic freedom has been undertaken in Refs.\cite{oz}.
  The gluon propagator satisfies an unsubtracted spectral representation
\be
\label{klg}
D(k^2, \alpha_{s},\mu)=
\frac{1}{\pi}\int\limits_0^{\infty} \frac{\rho_{v}(\sigma,\alpha_{s},\mu)}
{\sigma-k^2-i0}d\sigma.
\ee
For a limited number of flavors $N_{f}\leq 9 $
in the Landau gauge
the weight function $\rho_{v}$ satisfies the superconvergence relation
$
\int\limits_0^{\infty}\rho_{v}(\sigma,\alpha_{s},\mu) d\sigma= 0. $
It was argued in Refs.\cite{oz}
that For $N_{f}\le 9$ there is an renormalization invariant point
$K^{2}(\alpha_{s},\mu)$ such that $\rho_{v}$ is a negative measure for
$\sigma\geq K^2$.
We can start from formula (\ref{gs}). The solution for the
$\beta$-function is known
in APT \cite{shir2}.
But,   the anomalous dimension function
$\gamma_{v}$ is not determined within this method yet.
One possible way is to  assume  following  relation
\be
\label{rlt}
\gamma_{v,an}^{(n)}(\alpha_{s})/\beta_{an}^{(n)}(\alpha_{s})=
\gamma_{v}^{(n)}(\alpha_{s})/\beta^{(n)}(\alpha_{s}),
\ee
where    $ \beta^{(n)} $ and  $\gamma_{v}^{(n)}$ are standard perturbative
expressions  for the renormalization group functions
in the n-loop order,       and
$ {\beta}^{(n)}_{an} $    and ${\gamma}_{v,an}^{(n)}$ are  corresponding
quantities  in APT.
Then from formula (\ref{gs}), in the one-loop order, we may write
\be
\label{1lgs}
 D^{(1)}_{an}(-k^2)=-\frac{1}{k^2}\left(\frac{\bar\alpha_{an}^{(1)}(-k^2)}
 {\alpha_{s}}\right)^{\gamma_0/\beta_0}=
 -\frac{1}{k^2}c_{v}\left(\frac{1}{\ln(-k^2/{\Lambda}^2)}+
 \frac{1}{1+k^2/{\Lambda}^2}\right)^{\gamma_{0}/\beta_{0}},
   \ee
with    $c_{v}=(\frac{4\pi}{\beta_{0}\alpha_{s}})^{\gamma_{0}/\beta_{0}}$.
It is easy to convince that expression (\ref{1lgs}) has correct analytical
properties,
since, $\bar\alpha_{an}^{(1)}(-k^2) $ is analytic and it does not
vanish  in the finite part of the complex $k^2$-plane.
Thus, function (\ref{1lgs}) satisfies   the K\"{a}llen-Lehmann representation.
The corresponding weight function is given by
\be
\label{gsp}
\rho_{v}^{(1)}(\sigma)=\rho_{+}(\sigma)+\rho_{-}(\sigma), \ee where
 $\rho_{+}$ ( $\rho_{-}$) is positive (negative)
measure, $ \rho_{+}(\sigma)=c_{v}\pi\delta(\sigma)$ and
$$
\rho_{-}(\sigma)\equiv\bar\rho_{-}(\bar\sigma)=
-c_{v}\frac{1}{\bar\sigma{\Lambda}^2}[R(\bar\sigma]^{\frac{\gamma_{0}}
{\beta_{0}}}
\sin\left[\frac{\gamma_{0}}{\beta_{0}}\arccos
\left(\frac{1}{R(\bar\sigma)}\left(\frac{\ln\bar\sigma}{\ln^{2}\bar\sigma
+\pi^2}+\frac{1}{\bar\sigma+1}\right)
\right)\right],
$$
here  $\bar\sigma=\sigma/{\Lambda^{2}}$ and
$$ R(\bar\sigma)=\sqrt{\left(\frac{\ln\bar\sigma}{\ln^2\bar\sigma+\pi^2}+
\frac{1}{1+\bar\sigma}\right)^2
+\left(\frac{\pi}{\ln^2\bar\sigma+\pi^2}\right)^{2}}. $$
The superconvergence relation  in this case reads
$ \int\limits_{0}^{\infty}\rho_{-}(\sigma)d\sigma=-\pi c_{v}$.
For the RG invariant scale $K^2(g,\mu^2)$, which has been introduced in
Refs.\cite{oz},
the solution (\ref{1lgs}) predicts trivial value $K^2(g,\mu^2)=0$.
The weight function (\ref{gsp}) has integrable singularity at $\sigma=0$.

For the two-loop   gluon propagator similar procedure leads
\be
\label{2lg}
d^{(2)}_{an}(-k^2)=\bar d^{(2)}(\bar\alpha_{an}^{(2)}(-k^2))/\bar
d^{(2)}(\alpha_{s}),
\ee
where
$$
\bar d^{(2)}(\alpha)=(\alpha)^{\gamma_{0}/\beta_{0}}(1+\alpha\beta_{1}/
4\pi\beta_{0})^{-(\gamma_{0}/\beta_{0}-\gamma_{1}/\beta_{1})},
$$
it is evident that (\ref{2lg}) satisfies requirement of analyticity, since,
the equation $\bar\alpha_{an}^{(2)}(-k^2)=-4\pi\beta_{0}/\beta_{1}<0$
has no solutions  in the cut complex $k^2$-plane. The reason is that
$\bar\alpha^{(2)}_{an}(-k^2)$ itself satisfies the K\"{a}llen-Lehmann representation
with
positive
defined weight function.
One can assume  the relations similar of (\ref{rlt}) for the  quark and
the ghost fields too, and  by the way, write
"analytically improved" solutions for the corresponding
propagators.

Now we suggest different  method for restoring analyticity
and renormalization invariance in perturbation theory.
In this case we do not accept the relation  (\ref{rlt}). Instead, we start
from the  perturbative expression for the gluon propagator.
To the one-loop approximation, in the Landau gauge, we have
\be
\label{gpts}
d^{(1)}(-k^2/{\mu}^2,\alpha_{s})=\frac{1}{1+\frac{\alpha_{s}}{4\pi}\gamma_{0}
\ln(-\frac{k^2}{{\mu}^2})}=
\left(\frac{\beta_{0}}{\gamma_{0}a_{s}}\right)
\frac{1}{\ln(\frac{-k^2}{{\mu}^2})-\frac{\beta_{0}}{\gamma_{0}}\ln(\xi)} ,
\ee
here $\gamma_{0}$ is given by (\ref{gv}),
$a_{s}=\frac{\beta_{0}}{4\pi}\alpha_{s}$,
$\xi={\Lambda}^2/{\mu}^2$ and
 we have used standard perturbative definition of the $\Lambda$ parameter
$ {\Lambda}^2={\mu}^2e^{-\frac{1}{a_{s}}} $.
The "analitization procedure" for (\ref{gpts}) leads
\be
\label{apgp}
d^{(1)}_{an}(u,\alpha_{s})=\bar d^{(1)}_{an}(u,\alpha_{s})/
\bar d^{(1)}_{an}(1,\alpha_{s})
\ee
where
\be
\label{ex}
 \bar d^{(1)}_{an}(u,\alpha_{s}) = \left(\frac{1}{\ln u-
 \frac{\beta_{0}}{\gamma_{0}}\ln \xi}-
\frac{{\xi}^{\frac{\beta_{0}}{\gamma_{0}}}}
{u-{\xi}^{\frac{\beta_{0}}{\gamma_{0}}}}\right),
\ee
and $ u=-k^2/{\mu}^{2}$.
In order to preserve the normalization (see (\ref{gl})) we have introduced in
(\ref{apgp}) the necessary nonperturbative factor. Furthermore,
the invariant scale $\Lambda$ in  (\ref{apgp}) now is determined
according  APT, since, the procedure implies the replacement
$\Lambda_{pt}\rightarrow\Lambda_{an} $.
Thus, to first order according APT \cite{shir1}, we have the relation
$$
\label{scl}
a_{s}=\frac{\beta_{0}}{4\pi}\alpha_{s}=-\frac{1}{\ln \xi}
-\frac{\xi}{1-\xi}=\frac{1}{\phi}+\frac{1}{1-e^{\phi}}
$$
where $\phi=-\ln\xi$. Using (\ref{apgp})
we  find the anomalous dimension for the gluon field
\be
\label{ad1}
\gamma^{(1)}_{v}(\alpha_{s})\equiv\bar\gamma^{(1)}_{v}(\phi)=
\lim_{u\rightarrow 1} u\frac{\partial}{\partial u}
\ln d^{(1)}_{an}(u,\alpha_{s})=
\frac{-\frac{(\gamma_{0}/\beta_{0})^2}{{\phi}^2}
+\frac{e^{-(\beta_{0}/\gamma_{0} )\phi}}{(1-e^{-(\beta_{0}/\gamma_{0})\phi})^2}
}
{\frac{(\gamma_{0}/\beta_{0})}{{\phi}}
-\frac{e^{-(\beta_{0}/\gamma_{0} )\phi}}{1-e^{-(\beta_{0}/\gamma_{0})\phi}}}.
\ee
The expression (\ref{apgp}) does not obey renormalization invariance.
We can consider
(\ref{apgp}) as a known estimate and then use Eq~(\ref{gs})
to derive the improved
solution.
Taking into account   the identity
$$
\int_{\alpha_{s}}^{\bar\alpha_{s}(-k^2)}\frac{\gamma_{v}(x)}{\beta(x)}dx
=\int_{\ln({\mu}^2/{\Lambda}^2)}^{\ln(-k^2/{\Lambda}^2)}
\bar\gamma_{v}(\phi)d\phi,
$$
from Eqs. (\ref{gs}) and (\ref{ad1}) we obtain
\be
\label{sl1}
d^{(1)}_{im}(-k^2/{\mu}^2,\alpha_{s})=
exp\left(-\frac{\gamma_{0}}{\beta_{0}}
\int_{\frac{\beta_{0}}{\gamma_{0}}\ln({\mu}^2/{\Lambda}^2)}
^{\frac{\beta_{0}}{\gamma_{0}}\ln(-k^2/{\Lambda}^2) }
\frac{(e^t-1)^2-t^2e^t}{t(e^t-1)(e^t-1-t)}dt\right).
\ee
The integral in (\ref{sl1}) can be performed explicitly.  This yields
\be
\label{sl2}
d^{(1)}_{im}(-k^2/{\mu}^2,\alpha_{s})=
\bar d^{(1)}_{im}(-k^2/{\Lambda}^2)/
\bar d^{(1)}_{im}({\mu}^2/{\Lambda}^2),
\ee
where
\be
\label{sl3}
\bar d^{(1)}_{im}(-k^2/{\Lambda}^2)=
\left(\frac{1}{\frac{\beta_0}{\gamma_{0}}\ln(-k^2/{\Lambda}^2)}
-\frac{1}{(-k^2/{\Lambda}^2)^{\frac{\beta_0}{\gamma_0}}-1}
\right)^{(\gamma_0/\beta_0)}.
\ee
We see that function (\ref{sl2}) satisfies the K\"{a}llen-Lehmann analyticity
for $\gamma_{0}/\beta_{0} \ge 0.5$. The corresponding flavor condition
is $N_{f}\le3$. For $N_{f}>3$ the unphysical singularities appear in the first
Riemann sheet.
This limitation for $N_{f}$ turns out to be natural.
Since, the singularities in (\ref{sl3}) occurred at $|k^2/{\Lambda}^2|=1$ where
the number of active quarks are just three. For this reason we cannot
reject the solution (\ref{sl3}) using arguments of analyticity
\footnote{The author owes to D.V. Shirkov for drawing his attention to this
peculiarity of the solution (\ref{sl3}).}.

Thus, we need independent criterion to  select the suitable analytic solution.
Recently Such a criterion has been suggested  in Ref.\cite{aa},
where the principle of minimality
for the nonperturbative contributions in perturbative (ultraviolet) region
has been formulated.
Using this principle one can easily verify that the   solution
 (\ref{sl2}) is preferable. Indeed, (\ref{sl2}) predicts more rapid decrease
 of the nonperturbative contributions
in ultraviolet region  then the solution (\ref{1lgs})
provided  that $66/39\le\beta_{0}/\gamma_{0}\le 2$
for $0\le N_{f}\le 3$.\\
\bigskip

{\bf 4. Conclusion}\\

The structure of the   $\beta$-function   
has been analyzed
in APT.
Modified expansion
for the $\beta$-function is written as power series   in
$\alpha_{s}$ with coefficients which  depend on  the QCD scale parameter
$\Lambda$.
The coefficients  are regular 
at $\Lambda=0$.

We have found compact expression for the analytic two-loop effective coupling
constant in terms of the Lambert W function.
This expression is   convenient to
explore  analytical  properties of the solution.

The method of APT has been extended for the   gluon propagator in the
Landau gauge.
In this case, as we have seen, the nonperturbative ambiguity arises.
The RG and analyticity constraints alone are not sufficient to uniquely
determine solution for the gluon propagator.
It was shown that application of the principle of minimality
for the nonperturbative contributions in perturbative (ultraviolet) region
\cite{aa}
allows us to reduce the ambiguity.

The  procedure does not destroy renormalizability of the  theory.
The only effect  is that the renormalization constants receive
finite nonperturbative corrections. This was demonstrated by explicit
calculations
of the $\beta$-function and the anomalous dimension function of
the gluon field.\\
\bigskip

{\bf Acknowledgments}\\

The author is very much indebted to the Organizers  of the
International Conference\\ " QUARKS-98" at Suzdal, for an opportunity to
participate in the conference
and for their warm hospitality.
The author wish to thank D.V. Shirkov for recent generous hospitality in
Dubna and for very helpful and interesting conversations  which stimulated
the investigation described here.
It is a pleasure to acknowledge
B.A. Arbuzov, V.A. Matveev and A.N. Tavkhelidze
for valuable discussions on this topic,
R. Bantsuri, M.A. Eliashvili, R.N. Faustov, G. Jorjadze, A.L. Kataev, 
D.I. Kazakov,
S.V. Mikhailov, A.V. Nesterenko, A.A. Pivovarov,
I.L. Solovtsov, O.P. Solovtsova and V.I. Yasnov
for  useful discussions.\\

\end{document}